# REAL-TIME SIMULATIONS TO VALIDATE THE IMPACT OF M-SSSC DEVICES ON PROTECTION COORDINATION IN POWER SYSTEMS

| S. HINCAPIE* | J.A. CALDERON | C.E. BORDA |
| Smart Wires Inc. | ISA | Smart Wires Inc. |
| Colombia | Colombia | Colombia |
| sebastian.hincapie@smartwires.com | jacalderon@isa.com.co | carlos.borda@smartwires.com |

| A. DUQUE | P. MACEDO | J.P. GALLEGO |
| Smart Wires Inc. | Smart Wires Inc. | ISA TRANSELCA |
| Colombia | United States | Colombia |
| alejandro.duque@smartwires.com | pablo.macedo@smartwires.com | jpgallego@transelca.com.co |

***Abstract*** – This paper analyzes the impact of the integration of the Modular Static Synchronous Series Compensator (M-SSSC) solution that will be deployed in the Santa Marta 220 kV (Colombian substation) by using real-time simulations with Hardware in the Loop (HIL) technique, over the protection coordination schemes in the area of influence. The results show that the M-SSSC devices can be successfully coordinated with the protection schemes, mitigating overloads through the transmission lines during steady state and contingencies, while avoiding potential harmful interactions with the grid and, most importantly, with protection devices. This behavior is achieved by using overcurrent detection logics implemented on M-SSSC devices that trip fast-acting Silicon Controlled Rectifiers (SCRs) that cease the voltage injection in less than 5 ms.

## 1 INTRODUCTION

Power Systems are facing an energy transition around the increasing integration of renewable resources. To be able to transport the energy from this rapidly growing generation, it is necessary to build new infrastructure or optimize the existing grid. Due to environmental policies and a lack of physical space, the first alternative is increasingly becoming unfeasible, making technologies that optimize the existing network even more relevant. M-SSSC [1] has been globally used to actively control power flows in the existing grid by pushing power off congested lines and/or pulling power towards underutilized corridors [2]. This effect is achieved by injecting a series voltage in quadrature to the line current, thereby changing the net impedance and effectively optimizing the grid utilization. Colombia is no stranger to this challenge. Due to the high potential of solar and wind resources in the north of the country, the transmission network must be rapidly upgraded. Considering it will take more time to build the network compared to the time for the new generation to be ready to connect, M-SSSC provides temporary, and definitive solutions.

One of the key aspects when integrating M-SSSCs into the system is their impact on protections schemes in the area. Since M-SSSC's voltage injection has a direct impact on the equivalent reactance of the line, this could modify the response of distance relays. This paper describes the process and results carried out to analyze the impact of two M-SSSC deployments connected in the Santa Marta 220 kV substation in Colombia using real-time simulations. The main objective of this M-SSSC solution is to avoid thermal overloads of the Santa Marta - Guajira 220 kV and Santa Marta – Termocol 220 kV circuits in N-1 conditions. The methodology and scope of the tests were developed jointly by a multidisciplinary team of the utility (ISA) and the developer of the M-SSSC (Smart Wires). Tests not only achieved an adequate coverage of cases that ensures a safe integration of the M-SSSCs from the protection perspective, but also the definition of improvements to the existing functionalities and the development of a backup scheme provided by the protection relays at the substation.

## 2 M-SSSC SOLUTION IN THE COLOMBIAN POWER GRID

### 2.1 Operating Principle of M-SSSC

---

* Calle 7D# 43A – 40, Medellín, Colombia - sebastian.hincapie@smartwires.com

The assessed M-SSSC, injects a leading or lagging voltage in quadrature (shifted 90 degrees) with the line current, providing the functionality of a series reactor or series capacitor respectively. However, unlike conventional series capacitors or reactors, M-SSSCs can inject the voltage independently of the line current as shown in Fig 1 (left) thus controlling the effective line reactance as shown in Fig 1 (right).

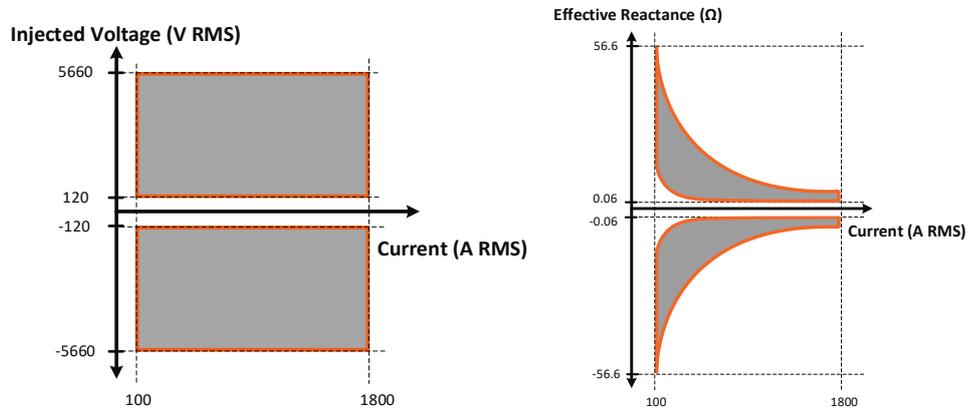

Fig 1.  M-SSSC Voltage Operating Range (left) and Reactance Operating Range (right)

An M-SSSC acts as a solid-state synchronous voltage source, consisting of a series of VSCs as shown in Fig 2. The H-Bridge of each converter uses IGBTs to inject a voltage directly into the network facility to maintain a desired reactance. This is achieved by sensing the line current through a current sensor to determine the correct voltage magnitude to inject. Typical use cases of M-SSSC include optimizing transmission margins by releasing spare capacity in the system, alleviating thermal overloads by re-directing power flows in the existing transmission lines, and accelerating renewable interconnection by solving bottlenecks [3]. M-SSSC can operate in several different control modes for these use cases, including Fixed Voltage and Fixed Reactance injection. Additionally, the M-SSSC can receive dynamic set points sent remotely from integrated control centers, selected based on external signals via SCADA protocols. This flexibility in its operating principles and control capabilities offers several benefits based on the action of "dispatching" line reactance.

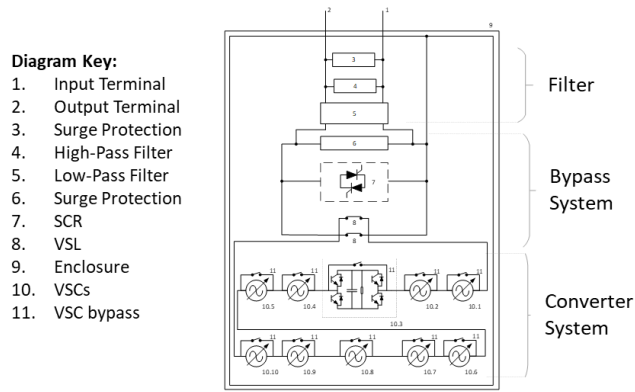

Fig 2.  M-SSSC Electrical Configuration

## 2.2  Primary components of M-SSSC

### 2.2.1  Filter
During system transients such as lightning surges, traveling waves can be induced into the M-SSSC. To mitigate the effects of said waves causing large voltage differentials across the M-SSSC, the filter system provides a low-impedance path for high-frequency components of line current while allowing currents around the system's fundamental frequency to pass through the bypass and converter systems.

### 2.2.2  Bypass System

The bypass system provides protection and control to the converters. The principal components of the bypass are the redundant normally-closed Vacuum Switch Link-VSLs, Silicon controlled rectifier-SCRs and a Surge Protection. The bypass system enables the rapid bypass of the converters during faults and lets operators switch the M-SSSC in series with the line.

**2.2.3   Converter System**
The converter system is in charge of injecting the voltage in series with the transmission line. In the case of the M-SSSC selected for this project, it is formed by 10 converters in series, each one consisting of an H-bridge capable of injecting up to 1 MVAr at rated current.

**2.3   M-SSSC expected behavior during faults**
The M-SSSC system incorporates an instantaneous Over-Current (OC) protection feature. The bypass is triggered if any phase current is larger than a user-selected threshold; thus, it removes the voltage injection from the system [4]. The bypass is composed by two redundant normally closed VSL and a fast-acting SCR branch that, primarily, conduct the current during grid faults. The bypass occurs within 1 ms from when a current threshold level is exceeded. Thus, the bypass feature minimizes the risk of interactions with existing protection schemes. After the device goes to bypass, it will remain in this state for at least 30 s after the fault is cleared to resume the injection.

M-SSSCs are inherently single-phase devices, and each phase monitors the magnitude and phase angle of the line current independently. An Interphase Balancing feature (IPB) can be enabled for asymmetrical faults, whereby devices on healthy phases bypass to avoid an unbalanced operation.

In addition to the Over-Current (OC) protection, there is a feature that allows the M-SSSC to resume the injection much faster compared to the normal OC protection. The Low-Overcurrent Ride-Through (LOR), if enabled, allows the SmartValve to temporarily bypass to fault currents below the hardware protection threshold and without the need of closing the VSLs, allowing it to return to Injection Mode faster than the reinjection time from the Over-Current protection. Besides the M-SSSC built-in LOR logic, and to consider all possible scenarios that could cause an unbalance condition in the system, a LOR backup logic was proposed and validated. This logic is shown in Fig 3 and consists of commanding the M-SSSC to activate the LOR response on the three phases of the M-SSSC when any of the conditions in the figure is met. This backup LOR logic is wired directly from the protection relay to the control and communication system of the M-SSSC. It will act alongside the built-in protection functions of the M-SSSC to conform a very robust system that ensures the correct behavior of the device and the protection relays of the area of influence of the project.

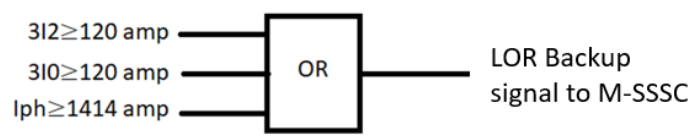

Fig 3. LOR backup logic

**2.4   M-SSSC at the Santa Marta 220 kV substation**
The Colombian Mining and Energy Planning Unit (UPME) included an expansion project associated with the installation of M-SSSCs in the Santa Marta 220 kV substation [5]. UPME considered the use of said M-SSSC solutions in fixed inductive reactance mode by identifying operation scenarios in which unacceptable overloads may occur in the Guajira - Santa Marta - Termocol 220 kV transmission corridor. The reactance setpoint values were selected to ensure the line current levels remain below the emergency overload limits for each circuit during N-1 conditions caused by the integration of generation plants that have been assigned firm energy supply requirements. Fig 4(left) shows the solution's influence area, located in the GCM operating sub-area of the Colombian National Interconnected System (SIN in Spanish).

UPME proposed a M-SSSC solution based on the installation of two M-SSSC deployments as follows:
- On the Santa Marta – Termocol 220 kV circuit: 15 M-SSSCs -5 per phase
- On the Santa Marta – Guajira 220 kV circuit: 9 M-SSSCs -3 per phase

The solution also takes advantage of the possibility of relocating M-SSSC units. Thus, it will have a second stage with the following reconfiguration:
- Relocation of the 15 M-SSSCs on the Santa Marta - Termocol 220 kV circuit to the Termoguajira - Termocol 220 kV circuit.

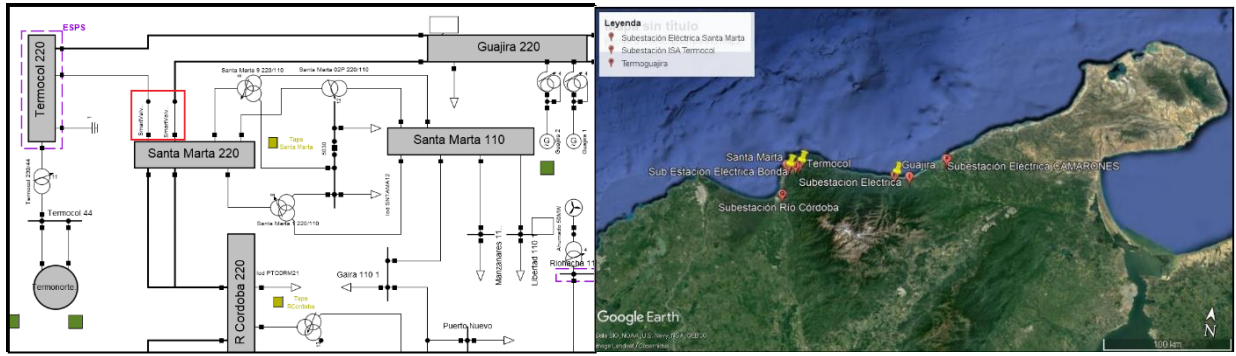

Fig 4 Santa Marta 220 kV M-SSSC solution (left) and Geographical location of M-SSSC system (right)

The Santa Marta 220 kV and Termocol 220 kV substations are geographically located in the city of Santa Marta, and Termoguajira 220 kV is geographically located in the municipality of Dibulla in the department of La Guajira. Fig 4 (right) shows the geographical location of the solution.

## 3 REAL TIME SIMULATION TESTS

Once the behavior of the M-SSSC devices in steady state conditions has been identified and their ability to mitigate potential overloads has been verified via system integration studies, it became relevant to validate the dynamic behavior of the real controls when facing system disturbances of different characteristics that could occur in the area of influence of the project by using real-time simulations [6]. The designed benchmark analyzes the so-called Dynamic Performance Tests (DPT). The results of these tests become relevant to validate the interaction of M-SSSC with the protection schemes of the area of analysis. This validation will help operators anticipate the system's behavior before the occurrence of typical and critical events in the power grid.

### 3.1 Control Hardware in the Loop (C-HIL)

The Hardware in the Loop (HIL) testing methodology is based on connecting a software model implemented in a Real Time Digital Simulation (RTDS) tool with actual physical hardware to observe the system behavior or response to a set of realistic conditions [6].

To test and verify the performance of M-SSSC system in the customer's power network, a Control Hardware in the Loop (C-HIL) test setup has been developed by utilizing Real Time Digital Simulator as shown in Fig 5. RTDS is an ideal tool for testing power electronics control devices in power systems. The purpose of the C-HIL is to avoid potential bugs associated with Firmware and Hardware before installing the device in the field.

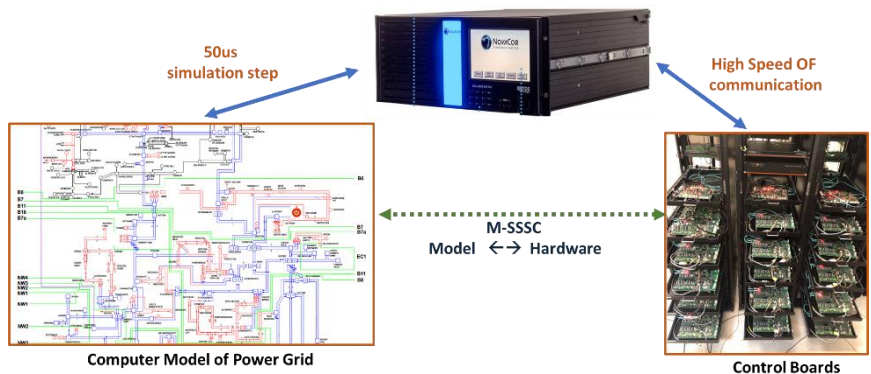

Fig 5. C-HIL overview for an M-SSSC solution

### 3.2 Hybrid deployment

To account for several M-SSSCs connected in series, a hybrid arrangement was implemented. One M-SSSC is studied in C-HIL, and the remaining M-SSSCs are represented as RSCAD models in a Software-In-Loop (SIL) arrangement. The bi-directional communication between C-HIL and SIL is done via RTDS GTNET cards [7] and M-SSSC's proprietary communication equipment.

For instance, Fig 6 shows a graphical representation of the Virtual deployment when it is located on the Santa Marta – Guajira 220 kV line, where the first M-SSSC corresponds to the C-HIL and the remaining two are models. The three of them are being coordinated by a Virtual Deployment setup that emulates the real performance of the actual deployment. On the other hand, in the Santa Marta – Termocol 220 kV the whole deployment is represented by one M-SSSC model scaled by 5 due to real time processing limitations. Likewise, when the C-HIL is located on Santa Marta – Termocol 220 kV line, there are 4 M-SSSC models in series to the HIL to complete the deployment, and there is one model scaled by 3 in Santa Marta – Guajira 220 kV.

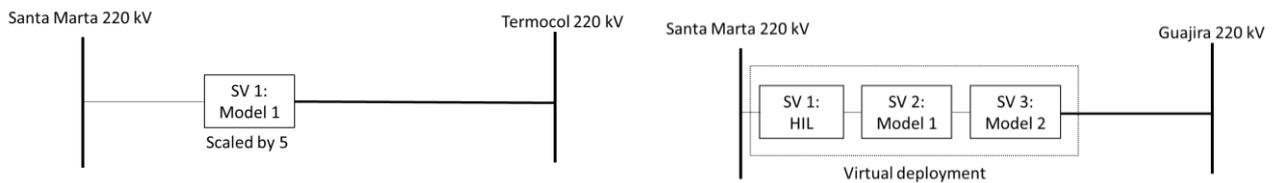

Fig 6. Virtual deployment

### 3.3 Protection relays in the loop (PIL)

Two SIEMENS 7LS87 protection relays were connected to the RTDS setup (Protection In the Loop - PIL) to validate the impact of the M-SSSC devices over their expected operation. The analog inputs of the relays were connected to the analog outputs of the RTDS by using electrical amplifiers, and the digital outputs of the relays go directly to the RTDS. The relays' settings files are the same to be used at the substation to ensure the exact same response of the protection devices connected in the lines where the M-SSSC deployments will be located.

### 3.4 Modelling of the network

As mentioned before, the project is located in the GCM operating sub-area of the Colombian National Interconnected System (SIN), then the whole subsystem was modeled to have the best representation of the grid. The official database managed by the Colombian Power System Operator (XM) was used to build the grid model in RSCAD, including elements such as lines, transformers, static shunt compensation, loads, substations, and generators with their controllers. Once the model was built, a comprehensive validation was carried out against the official model, emphasizing in the response considering different faults in the grid.

### 3.5 Relay and M-SSSC models

In addition to the elements modeled and described in the section above, all the Transmission Line Protection Relays of the area of analysis were modeled to complement the PIL setup. The relay model has multiple logics that emulate the real behavior of the real devices such as distance loops, directional detection, signals processing, trip logic, auto-reclosing logic, evolving fault logic, ANSI 21 detection, and more.

Similarly, since the C-HIL setup can represent up to three M-SSSC single-phase devices, a detailed M-SSSC EMT model was developed using RSCAD's C-builder platform to represent all its features. The model has been validated against the C-HIL and as mentioned before both C-HIL and SIL are connected in series to represent the whole deployment.

## 4 DPT METHODOLOGY AND RESULTS

Considering the whole setup described in the previous sections, more than 20 simulations and 6 different types of faults were carried out to validate the impact of the M-SSSC devices on the protection coordination of the area of analysis. There are three relevant cases to demonstrate not only the effectiveness of the M-SSSC technology to avoid overloads but also the negligible impact that it has on the protection coordination schemes.

Next cases consider a three-phase fault at 50% of the Santa Marta – Guajira 220 line with a fault impedance of 5 Ohm and a duration of 1.5 s.

**4.1** Case 1: GCM sub-area without the integration of the M-SSSC devices

Fig 7 shows relay and network performance considering the fault mentioned above and no injection of the M-SSSC deployments. As the fault is applied in Santa Marta – Guajira 220 kV, the line protections cleared the fault by opening the circuit breakers of the line, increasing the current in the Santa Marta – Termocol 220 kV. Around 900 ms after, the relays tried to reclose the line without success leaving the circuit breakers open. At the end of the simulation, the current through the Santa Marta – Termocol 220 kV line is over 0.8 kA which exceeds the thermal loading limit of the conductor (787 A). This condition will effectively lead to violation of Colombia's grid code or to a possible N-1-1 contingency that increases the risk of a total blackout of the GCM operating area.

**4.2** Case 2: GCM sub-area with the integration of the M-SSSC devices but without the bypass features (LOR or/and OC).

Fig 8 shows the performance of the network with the injection of the M-SSSC deployments but without the bypass features. In this case, like the previous case, the Santa Marta – Guajira 220 kV is tripped by the line protection and the M-SSSC deployment goes to bypass due to the current through the line is zero. On the other hand, the voltage injection of the M-SSSC deployment in the Santa Marta -Termocol 220 kV is behaving in a way that is affecting the behavior of the protection schemes. Due to the natural dynamic response of the system after a fault, the PLL of the M-SSSC is struggling to follow the reference of the current, causing no ideal performance of the device and the system. This case demonstrates that if there is not an adequate response of the M-SSSC devices during disturbances, it could affect the whole behavior of the system including the protection schemes and again, implying risks of a total blackout of the GCM operating area.

**Santa Marta – Termocol 220 kV**           **Santa Marta – Guajira 220 kV**

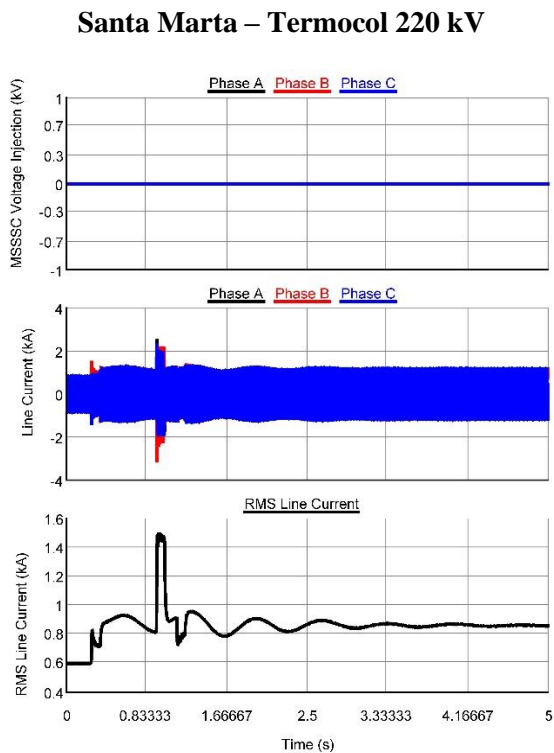
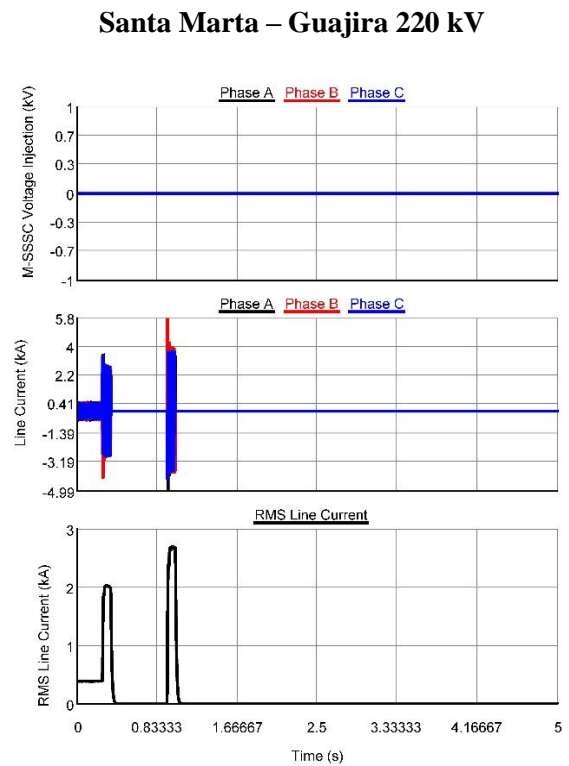

Fig 7-(a). Case 1. Santa Marta – Termocol 220 kV variables

Fig 7-(b). Case 1. Santa Marta – Guajira 220 kV variables

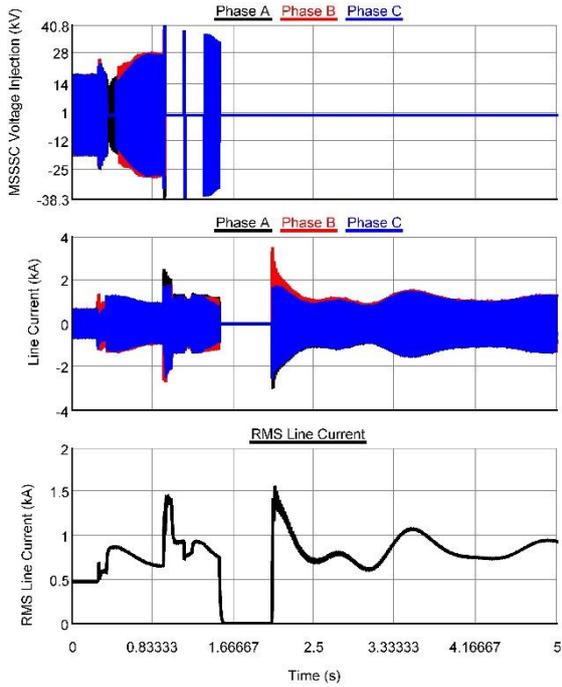

Fig 8.-(a). Case 2. Santa Marta – Termocol 220 kV variables

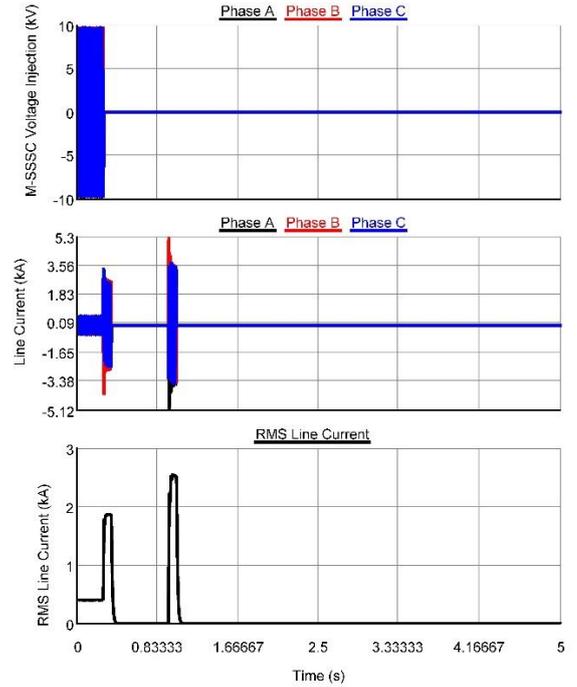

Fig 8-(b). Case 2. Santa Marta – Guajira 220 kV variables

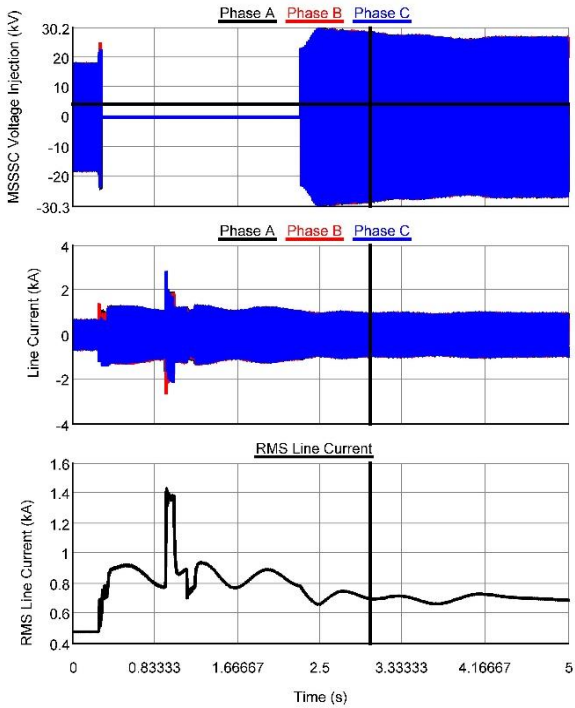

Fig 9.-(a). Case 3. Santa Marta – Termocol 220 kV variables

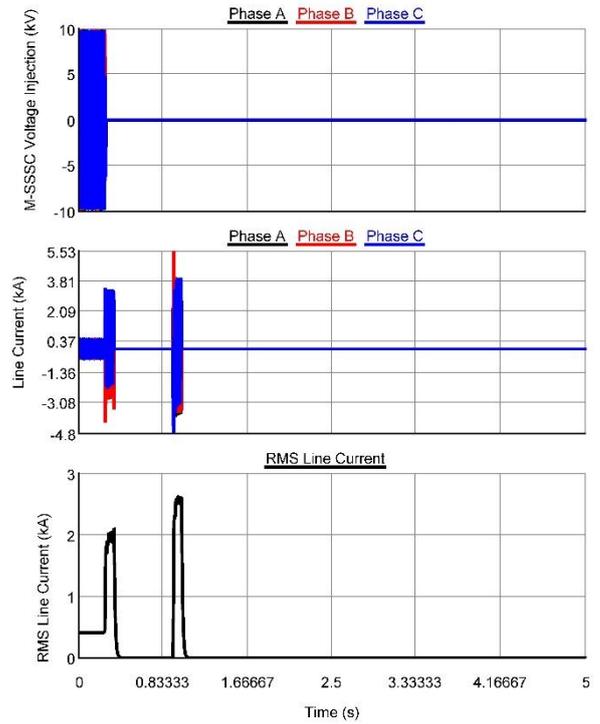

Fig 9-(b). Case 3. Santa Marta – Guajira 220 kV variables

**4.3** Case 3: GCM sub-area with the integration of the M-SSSC devices and with the bypass features (LOR or/and OC).

Fig 9 shows the third and last case, where the M-SSSC deployments are considered, as well as the bypass features. The relays in Santa Marta - Guajira 220 kV clear the fault and try to reclose the circuit breakers without success. In this case, the M-SSSC deployment on the faulty line goes to bypass due to the activation of the OC logic (current on Phase B higher than 3.8 kA in one of the phases), and then it remains in this state for at least 30 s. The M-SSSC in the healthy line goes to bypass due to the activation of the LOR logic (current higher than 1.1414 kA) and remains in this state for 1 s after the last LOR signal triggered (right after the reclosure on fault). As the figure shows, the fact that the M-SSSC deployment goes to bypass allows the desired recovery of the grid without any sort of harmful interaction with the protection schemes or the power system. In addition, the RMS current at the end of the simulation, when the fault transient finishes and the M-SSSC deployment is injecting, is less than 700 A, lower than the thermal loading limit of the conductor (787 A), demonstrating the effectiveness of the M-SSSC devices to control overloads in the GCM area.

Considering the previous cases and all the others that were run during this analysis, it was observed the importance of using the bypass logic and the LOR feature to guarantee an adequate operation of the M-SSSC devices during faults. In addition, in order to have a balanced operation of the system, it is recommended that the M-SSSC deployments have a balanced behavior, and here is where the IPB feature described in 2.3 as well as the three-phase LOR logic become relevant, due to the coordination between phases when asymmetrical faults occur in the system.

# 5 CONCLUSION

This work demonstrates that the M-SSSC devices that will be connected in the Santa Marta 220 kV substation (located in Colombia) can effectively control the power flow through the lines and mitigate overloads in steady state and during contingencies. In addition, the results show that is crucial to have an adequate response of the M-SSSC devices during faults to guarantee the desired performance of the grid and the protection schemes of the area of influence, and this performance is achieved, thanks to the fast-acting Silicon Controlled Rectifiers (SCRs) and the correct parametrization and coordination of the bypass logics integrated into the M-SSSC devices that allow a fast transition (less than 5 ms) from injection to monitoring mode avoiding harmful interactions between the M-SSSC devices, the protection schemes, and the grid. Finally, more than 20 simulations and 6 different types of faults, using real-time simulation, real hardware, and validated models against the real behavior of the actual devices, reveal the correct coordination and performance of the M-SSSC devices, that detected and operated accordingly considering different current conditions in the grid by discerning in which cases it must go to definitive bypass, LOR or continue injecting.